\documentclass[superscriptaddress,
longbibliography,
amsmath,amssymb,preprint,
prf]{revtex4-1}
\usepackage{graphicx}
\usepackage{epstopdf, epsfig}
\usepackage{graphicx}
\usepackage{dcolumn}
\usepackage{bm}
\usepackage{epsfig} 
\usepackage{amsmath}
\usepackage{float}
\usepackage[utf8]{inputenc}
\usepackage[T1]{fontenc}
\usepackage{nomencl}
\usepackage{CJKutf8}
\usepackage{xcolor}
\usepackage{subcaption}
\usepackage{comment}
\usepackage{makecell}
\usepackage{soul}
\usepackage[ruled,linesnumbered]{algorithm2e}

\makenomenclature
\begin{document}

\title{Simulation and analytical modeling of high-speed droplet impact onto a surface}

\author{Yanchao Liu (\begin{CJK*}{UTF8}{gbsn}刘雁超\end{CJK*})}
\email{wild_goose_liu@hotmail.com}
\affiliation{Institute of Aerospace Thermodynamics, University of Stuttgart,
Pfaffenwaldring 31, 70569 Stuttgart, Germany}

\author{Xu Chu (\begin{CJK*}{UTF8}{gbsn}初旭\end{CJK*})}
\affiliation{Cluster of Excellence SimTech (SimTech), University of Stuttgart,
Pfaffenwaldring 5a, 70569 Stuttgart, Germany}

\author{Guang Yang (\begin{CJK*}{UTF8}{gbsn}杨光\end{CJK*})}
\affiliation{Institute of Refrigeration and Cryogenics, Shanghai Jiao Tong University, 200240 Shanghai, China}

\author{Bernhard Weigand}
\affiliation{Institute of Aerospace Thermodynamics, University of Stuttgart,
Pfaffenwaldring 31, 70569 Stuttgart, Germany}



\begin{abstract}
The fluid dynamics of liquid droplet impact on surfaces hold significant relevance to various industrial applications. However, high impact velocities introduce compressible effects, leading to material erosion. A gap in understanding and modeling these effects has motivated this study. We simulated droplet impacts on surfaces and proposed a new analytical model for impact pressure and droplet turning line, targeting at predictions for enhanced cavitation. The highly compressed liquid behind the droplet expands sideways, causing lateral jetting. As the droplet encounters a shock wave, it reflects as a rarefaction wave, leading to low-pressure zones within the droplet. These zones converge at the droplet's center, causing cavitation, which, upon collapse, induces another shock wave, contributing to erosion. Using the well-established model for the low-velocity impact shows a significant discrepancy. Hence, an analytical model for the turning line radius is introduced, incorporating the lateral jetting's characteristic length scale. Comparing our model with existing ones, our new model exhibits superior predictive accuracy.

\end{abstract}





\maketitle

\section{Introduction}
\label{Intro}
The fluid dynamics of liquid droplet impact on surfaces hold significant relevance to various industrial and technological applications, such as spray cooling, ink-jet printing, rainfall, fuel atomization, and spray cleaning. Over the past several decades, investigations have been mainly focused on low velocity droplet impact, in which the compressible effects of the liquid are considered negligible. Nonetheless, in certain applications such as a high-fogging system in a gas turbine, a steam turbine, flight vehicles through rain or a medical inhaler with a liquid jetting nozzle and high-speed liquid jets in cleaning and cutting operations, the impact velocity of the droplet is relative high. In these context, the compressiblility effects of the fluid cannot be neglected and is instrumental in causing material erosion \citep{AHMAD20091605,FIELD2012154,BURSONTHOMAS2019507,IBRAHIM2022110312}. Despite the clear implications, research in this area, especially concerning high velocity impacts onto structured surfaces and their corresponding modeling, remains largely insufficient.

Theoretical analysis of high-velocity liquid droplet impingement dates back to \citet{bowden1964brittle}, where they showed that when a droplet impacts onto a surface at a high-velocity, shock waves may be generated inside the droplet. \citet{Heymann1969} estimated the maximum impact pressure using a two-dimensional approach. It was shown that the pressure in the contact area is not uniformly distributed and the highest pressure is located just behind the contact line.
\citet{Lesser1981} derived analytical solutions for the pressure dynamics in an impacting liquid drop against both rigid and elastic targets. They predicted that the pressure behind the contact line reaches its highest value when the shock wave detached from the contact line. In \citet{Haller2003}, an analytical exploration of the wave structure at the contact line region during a high-velocity liquid droplet impact has been carried out. They resolved the anomaly associated with the single shock wave assumption, underscoring the relevance of the proposed double wave structure as a more accurate representation of the physical phenomena. \citet{LI20081526} used a non-linear wave model to investigate the coupled transient pressure and stress fields in the liquid drop and metal substrate. Through both analytical and numerical solutions, it provided critical insights into the phenomena of continuous and pulsant impacts on rigid and elastic substrates.

Experimental investigations of high-velocity droplet impact are very challenging and limited owning to the complexity and small time scale of the problem. Mostly, the cameras and the resolution are not fast and sufficient enough to capture the detailed flow features of the impact. Among them, \citet{camus1971study} used a single-shot schlieren approach to capture the shock wave generation, propagation and reflection inside a droplet with an impact velocity of 70~m/s. \citet{Field1989} investigated two-dimensional drops, demonstrating the role of target compliance in delaying the onset of jetting, and defining conditions for the overtaking of the shock envelope by the contact periphery and subsequent jet appearance. \citet{FIELD2012154} presented high-speed photographic evidence of cavity formation and shock propagation in impacted liquids, showing that the cavity collapse process can indeed account for the observed lower threshold velocities. In \citet{NYKTERI2020109225}, droplet fragmentation after high-velocity impact has been represented experimentally.

With the improvement of the computational power and numerical methods, numerical simulation became an effective tool to investigate the high velocity droplet impact in recent years. \citet{Haller2002} simulated a water droplet with an impact velocity of 500~m/s. They compared the computationally obtained jetting inception times with analytic results and showed that jetting inception times are significantly enhanced when radial motion of the liquid within the compressed area is factored in. \citet{CHIZHOV20041391} investigated a high-velocity liquid nitrogen drop impact on a hot rigid wall, highlighting a particularly thin layer of fluid undergoing rapid heating and phase transition with the majority of the liquid remaining cold. In \citet{wu_xiang_wang_2018} and \citet{wu_liu_wang_2021}, the impact induced cavitation has been simulated and the shock wave generated by cavitation collapsing has been evaluated. The effects of impact velocity and surface curvature have been considered. \citet{kondo2019} focused on wall shear flow and water hammer effects following high-speed droplet impact onto dry and wet rigid surfaces. By simulating hydrodynamic forces on assumed wall-attached particles and comparing these forces to van der Waals type adhesion, the research presents a simple criterion for particle removal. \citet{NYKTERI2020109225} proposed a diffuse interface approach for unresolved liquid structures, showing the advantages of the new approach on evaluating the droplet fragmentation after impact onto the target with high-velocity. \citet{Marzbali2020} quantified the impact pressure of compressible liquid droplets on rigid substrates and liquid films and improved the correlations for maximum impact pressure. \citet{fluids6020078} applied a smoothing procedure to the interface treatment, demonstrating an improvement in the reduction of spurious currents in simulations of high-velocity droplet impact.

Despite previous studies, high-velocity droplet impact onto a structured surface, such as a pillar, has not yet been investigated. Furthermore, while \citet{cheng2022} summarized the modeling of parameters such as drop impact pressure, turning line radius, impact forces, and stress distributions for low velocity impacts, where compressibility effects are neglected, similar modeling for high-velocity droplet impacts remains unexplored. In the present study, we simulate a droplet impact onto cylindrical pillars of varying diameters and propose correlations for the peak impact pressure and the turning line of the droplet, aiming to predict pillar diameters in relation to a possible cavitation enhancement.

\section{Numerical approach}
\subsection{Governing equations}
To model the two phase flow, the Volume of Fluid (VoF) method \citep{hirt1981volume,liu2021simulation,liu2023large,yang2020droplet} is employed. In the VoF method, a scalar field $\alpha \in [0,1]$ is introduced to indicate the volume fraction of the liquid phase. By calculating the fluid properties such as density $\rho$ and viscosity $\mu$ as
\begin{equation}
    \label{rhoDef}
    \rho = \rho_l \alpha + \rho_g (1-\alpha)
\end{equation}
and
\begin{equation}
    \label{muDef}
    \mu = \mu_l \alpha + \mu_g (1-\alpha),
\end{equation}
the conservation of mass, momentum and energy for both phases can be described by
\begin{equation}
\label{massConserv}
    \frac{\partial \rho}{\partial t} + \nabla \cdot \left( \rho \mathbf{U} \right) = 0,
\end{equation}

\begin{equation}
    \frac{\partial (\rho \mathbf{U})}{\partial t} + \nabla \cdot \left( \rho \mathbf{U} \mathbf{U} \right) =
    - \nabla p + \nabla \cdot \pmb{\tau} + \mathbf{g} \cdot \mathbf{h} \nabla \rho +\mathbf{F}_{\sigma},
\end{equation}

\begin{equation}
\label{eq:energyConserv}
 \begin{aligned}
     \frac{\partial (\rho T)}{\partial t} + \nabla \cdot \left( \rho \mathbf{U} T \right) -\nabla \cdot \left( \alpha_{eff} \nabla T \right) =
     \& -\left( \frac{\alpha}{c_{v,l}} + \frac{1-\alpha}{c_{v,g}} \right)\cdot \\
     \& \left( \frac{\partial (\rho K)}{\partial t} + \nabla \cdot \left( \rho \mathbf{U} K \right) +\nabla\cdot\left( p \mathbf{U} \right) \right),
 \end{aligned}
\end{equation}
where $t$, $\mathbf{U}$, $p$, $\pmb{\tau}$, $\mathbf{g}$, $\mathbf{h}$, $T$, $\alpha_{eff}$ and $c_v$ represent the time, velocity vector, pressure, viscous stress tensor, gravity vector, position vector, temperature, effective thermal diffusivity and specific heat capacity at constant volume, respectively. The subscript $l$ and $g$ indicate the liquid and gas phases. $K$ denotes the specific kinetic energy, defined as $K = |\mathbf{U}|^2/2$. $\mathbf{F}_{\sigma}$ is the surface tension force.
The gas-liquid interface is captured by the advection of the volume fraction field
\begin{equation}
    \label{vofadv}
    \frac{\partial \alpha}{\partial t} + \nabla\cdot\left( \alpha \mathbf{U} \right) =
    \alpha(1-\alpha)\left( \frac{1}{\rho_g} \frac{\mathrm{D}\rho_g}{\mathrm{D} t} - \frac{1}{\rho_l} \frac{\mathrm{D}\rho_l}{\mathrm{D} t} \right) + \alpha \nabla \cdot \mathbf{U}.
\end{equation}
The derivation of this equation can be found in Appendix~\ref{sec:VoFeq}. 

To close the equations systems (\ref{massConserv}-\ref{vofadv}), equations of state for the gas and the liquid phase are needed. The thermodynamic state of the gas phase is described by the ideal gas law
\begin{equation}
    p_g = \rho_g R T_g,
\end{equation}
where $R$ is the specific gas constant that is set to 287~J/(kg$\cdot$K) for air. To account for the compressibility of the liquid phase, the Tait's power law equation of state \citep{tait1965report} is employed, which is the isentropic form of the stiffened gas equation of state defined as
\begin{equation}
\label{eq:EOS}
    \frac{p+B}{p_0+B}= \left( \frac{\rho_l}{\rho_{l_0}} \right)^N,
\end{equation}
where $p_0$ and $\rho_{l_0}$ are the pressure and the density at the reference state, respectively. Here we use water at ambient conditions as reference with $p_0=0.1$~MPa and $\rho_{l_0}=1000$~kg/m³. $B$ and $N$ are two constants equal to $B=300$~MPa and $N=7.415$.

The surface tension is calculated by the Continuum Surface Force model \cite{BRACKBILL1992335}, where the surface tension is represented as a body force. The surface tension force at any point in the domain is given by
\begin{equation}
    \mathbf{F}_{\sigma} = \sigma \kappa (\alpha) \nabla \alpha,
\end{equation}
where 
\begin{equation}
    \kappa(\alpha) = - \nabla\cdot (\,\mathbf{n}\cdot\mathbf{S}_f)\,
\end{equation}
is the surface curvature. $\mathbf{n}$ denotes the unit normal vector of the liquid-gas interface.  $\mathbf{S}_f$ denotes the outward-pointing vector of the cell face.

\subsection{Numerical methodology}
The finite volume method is used to discretize the equation system Eq.~(\ref{massConserv}-\ref{vofadv}). In Eq.~(\ref{vofadv}), since the geometric information of the interface is not contained in the volume fraction field, discretization of the advection term $\nabla\cdot\left( \alpha \mathbf{U}\right)$ leads to numerical diffusion that can make a sharp interface between two fluids appear artificially smeared over several grid cells. In order to accurately calculate the flux of the advection term as well as the interface curvature, the interface reconstruction method is applied. In the present study, the MPLIC (Multicut Piecewise-Linear Interface Calculation) interface reconstruction algorithm is employed, which performs a topological face-edge-face walk to produce multiple splits of a cell. The detailed description of the MPLIC algorithm is given in Section~\ref{sec:MPLIC}. To ensure the boundedness of the volume fraction, the multi-dimensional limiter for the explicit solution (MULES) algorithm, which is based on the flux-corrected transport method \cite{BORIS197338}, is used to limit the high order flux of Eq.~(\ref{vofadv}).
The advection terms in Eq.~(\ref{massConserv}-\ref{eq:energyConserv}) are discretize by a second order weighted essentially non-oscillatory (WENO) scheme implemented by \citet{GARTNER2020100611}. A first order implicit Euler scheme is utilized for time marching. The gradient is calculated by the Gauss linear scheme with central differencing.

\subsection{Description of the interface reconstruction algorithm MPLIC}
\label{sec:MPLIC}
Instead of approximating the interface inside a cell with a plane, this algorithm approximates the cross-section of the interface with the cell face using straight lines. Thus, the algorithm can be applied to general polyhedral meshes. The procedure largely follows the algorithm proposed by \citet{Roenby2016}, but with some improvements. In this section, the algorithm MPLIC will be briefly described.

In this approach, the volume fraction is interpolated to the cell vertices. To perform the face cut, an iso-value for the cell edges is calculated. With this value, the cell faces are cut by either single or multiple lines. The original algorithm proposed by \citet{Roenby2016} was reported to be less accurate, as mentioned in \cite{SCHEUFLER20191}. As a result, an additional correction was introduced. If the volume of the sub-cell deviates from the cell volume fraction by more than 10\%, the cell is decomposed into tetrahedrons. The face cut is then performed on these tetrahedrons to calculate the flux across the cell face. An overview of the algorithm is provided in Algorithm~\ref{alg:mplic}.
\RestyleAlgo{ruled}
\SetKwComment{Comment}{/* }{ */}
\begin{algorithm}
\caption{MPLIC algorithm}\label{alg:mplic}
\textbf{start}\\
create a list of cells that need to be cut, if cell volume fraction satisfies $\epsilon<\alpha<1-\epsilon$, where $\epsilon$ is a tolerance, which is set to $10^{-6}$.\\
\For{celli in the list}{
  Interpolate velocity and volume fraction to vertices of the cell using the surrounding cell volume fractions.\\
  Utilize the vertex values and interpolated values to cut the cell four times using the cut-face method (see Section~\ref{sec:face-cut} for details). Use these vertex values and calculated sub-cell volumes to build a cubic polynomial (see Section~\ref{sec:cubic} for details).\\
  Find the root of the cubic polynomial using the cell volume fraction to get the iso-value $\alpha_{viso}$ for the face cut.\\
  Use $\alpha_{viso}$ to cut the cell and calculate the sub-cell volume $V_s$ (see Section~\ref{sec:face-cut} for details).\\
  \If{1-mag($\alpha_{celli}$/$V_s$)>0.1}{
    decompose the cell into tetrahedrons.\\
    repeat steps 4-6 and cut each tetrahedron
  }
  Calculate the flux across the cell face (see Section~\ref{sec:face-cut} for details).
}
\end{algorithm}
\subsubsection{Method to map the vertex volume fraction values to the sub-cell volume}
\label{sec:cubic}
To cut the face, an iso-value of the volume fraction on the face is required. Since the geometry of the cell is unknown, deriving this iso-value directly from the cell volume fraction is non-trivial. In the current algorithm, a cubic polynomial is employed to approximate the geometric correlation between these two values. Initially, the vertex values are sorted as $\alpha_{v1},...,\alpha_{vN}$. Starting from the median value, two vertex values, $\alpha_{va}$ and $\alpha_{vd}$, are identified such that $\alpha_{va}<\alpha_{celli}<\alpha_{vd}$. Two interior values are then interpolated: $\alpha_{vb} = \alpha_{va} + (\alpha_{vd}-\alpha_{va})/3$ and $\alpha_{vc} = \alpha_{va} + 2(\alpha_{vd}-\alpha_{va})/3$. After calculating the sub-cell volumes $V_{sa},V_{sb},V_{sc},V_{sd}$ from these four values $\alpha_{va},\alpha_{vb},\alpha_{vc},\alpha_{vd}$ using face-cut method, we can derive the four coefficients of the cubic polynomial $V_s(\alpha_v)=a\alpha_v^3+b\alpha_v^2+c\alpha_v+d$. By substituting the cell volume fraction $\alpha_{celli}$ into the cubic polynomial and finding its root with respect to $\alpha_{celli}=a\alpha_v^3+b\alpha_v^2+c\alpha_v+d$, the iso-value $\alpha_{viso}$ for the face-cutting can be obtained. Details about the root-finding procedure can be found in \citet{Roenby2016}.
\subsubsection{The face-cut method}
\label{sec:face-cut}
This step aims to approximate the cross-section of the interface and the cell face using lines. Given the target value $\alpha_{vt}$, we can determine the corresponding point on an edge through linear interpolation. If only two such points are present on a face, the face is cut by a single line. The flux is then computed using the area of the resulting sub-face. However, if more than two points are present, the face is divided by lines connecting points on two adjacent edges. The flux across the cell face is subsequently calculated by decomposing the sub-face into triangles:
\begin{equation}
    \phi_{\alpha,f} = \sum_{N_t} \frac{1}{3}(\mathbf{u}_{t,v1}+\mathbf{u}_{t,v2}+\mathbf{u}_{t,v3})\cdot\mathbf{S}_t,
\end{equation}
where $N_t$ is the number of triangles contained in the sub-face. $\mathbf{u}_{t,v1,2,3}$ represent velocity values at the three vertices of the triangle. $\mathbf{S}_t$ is the area vector of the triangle. The volume of the sub-cell is approximated by decomposing the sub-cell into tetrahedrons. All tetrahedrons share te same vertex $\overline{\mathbf{x}}_f$ determined by averaging all vertices of the sub-cell. The volume of the sub-cell is expressed as:
\begin{equation}
    V_s = \sum_{N_f}\frac{1}{3}|(\mathbf{x}_f-\overline{\mathbf{x}}_f)\cdot \mathbf{S}_f|,
\end{equation}
where $N_f$ is the number of faces of the sub-cell, $\mathbf{x}_f$ and $\mathbf{S}_f$ are the face center and area vector of the sub-cell face, respectively. It is evident that with the present face-cut method, the calculated interface is not necessary planer. The face area vector of the interface is thus determined by averaging its triangular decomposition:
\begin{equation}
    \mathbf{S}_v = \sum^{N_v}_{k=1}\mathbf{S}_{v,k},
\end{equation}
where $N_v$ denotes the number of triangles contained in the polygonal interface and $\mathbf{n}_{a,k}$ is the area vector of the triangle calculated by
\begin{equation}
    \mathbf{S}_{v,k} = \frac{1}{2}(\mathbf{x}_{k+1}-\mathbf{x}_k)\times(\overline{\mathbf{x}}-\mathbf{x}_k)\textrm{ with } \overline{\mathbf{x}} = \frac{1}{N_v}\sum^{N_v}_{k=1}\mathbf{x}_k,
\end{equation}
where $\mathbf{x}_{Nv+1}=\mathbf{x}_{1}$. The normal vector and the center point of the interface are subsequently calculated as
\begin{equation}
    \mathbf{n} = \frac{\mathbf{S}_v}{|\mathbf{S}_v|} \textrm{ and } \mathbf{x} = \sum^{N_v}_{k=1}\frac{|\mathbf{S}_{v,k}|}{|\mathbf{S}_v|}\frac{\mathbf{x}_{k}+\mathbf{x}_{k+1}+\overline{\mathbf{x}}}{3}.
\end{equation}

\subsection{Validation of the numerical approach}
\label{sec:valid}
\begin{table}
\begin{tabular}{|c|c|c|c|}
\hline
Liquid 	& Density,$\rho$(kg$\cdot\mathrm{m}^{-3}$)	& Viscosity, $\mu$(Pa$\cdot\mathrm{s}$) & Surface tension, $\sigma$(N$\cdot\mathrm{m}^{-1}$)\\
\hline
Isopropanol   &   781.5\%   	&  0.00204		& 0.02092\\
\hline
\end{tabular}
\caption{Physical properties of isopropanol at 298.15~K}
\label{tab:prop_isop}
\end{table}

\begin{figure}[h]
\centering
    \begin{subfigure}[b]{0.48\textwidth}
        \centering
        \includegraphics[height=3.4in]{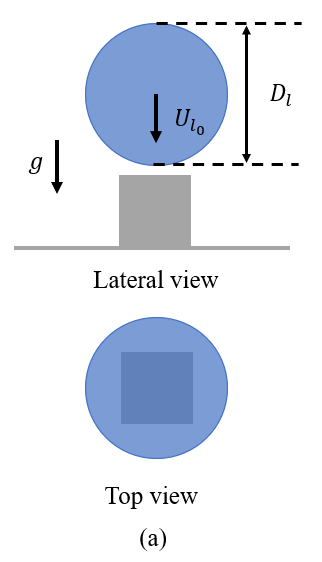}
    \end{subfigure}
    \begin{subfigure}[b]{0.48\textwidth}
        \centering
        \includegraphics[height=4.6in]{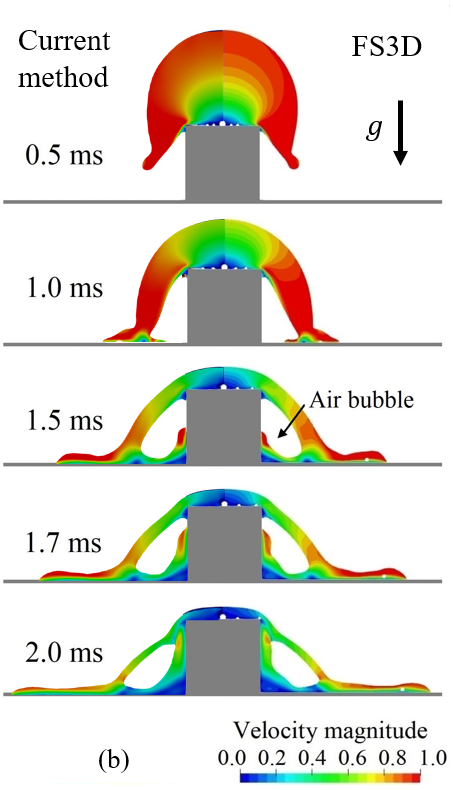}
    \end{subfigure}
\caption{Geometry and results for the validation: (a) Schematic depiction of a droplet impacting onto a cubic pillar; (b) comparison of normalised velocity field for a slice across the pillar centre and the centre of the pillar edges. Left: simulation results using current numerical approach. Right: simulation results from FS3D.}
\label{fig:valid}
\end{figure}

Before conducting simulations of high-velocity droplet impacts, the aforementioned numerical approach was validated against the DNS code FS3D \cite{fs3d}. FS3D is an incompressible VoF solver that employs the piecewise linear interface calculation (PLIC) method for multiphase flows. In the study by \citet{Ren2021}, the impact of a droplet onto a cubic pillar was simulated using FS3D and subsequently compared with experimental data, yielding a favorable agreement. As a result, the case of a central impact from their study was replicated using our current numerical approach. A schematic representation of this problem can be seen in Fig.~\ref{fig:valid}a. The scenario depicts a droplet impacting the exact center of a cubic pillar. The droplet has a diameter of $D_l= 2$~mm, and its impact velocity is $U_{l_0}=1.46$~m/s. The pillar measures $1\times1\times1$~mm. The liquid used in this simulation is isopropanol at a temperature of 298.15K. The physical properties of isopropanol are provided in Tab.\ref{tab:prop_isop}. A comparison of the normalized velocity field ($U/U_{l_0}$) between the two numerical approaches, for a slice through the pillar's center and the center of the pillar's edges, is illustrated in Fig.\ref{fig:valid}b by a side to side comparison between the velocity prediction of FS3D and the present method. The results indicate that our current numerical approach yields results closely aligned with those from FS3D. 

\section{Problem description}
\subsection{Simulation setup}
\begin{figure}[h]
\centering
\includegraphics[scale=0.5]{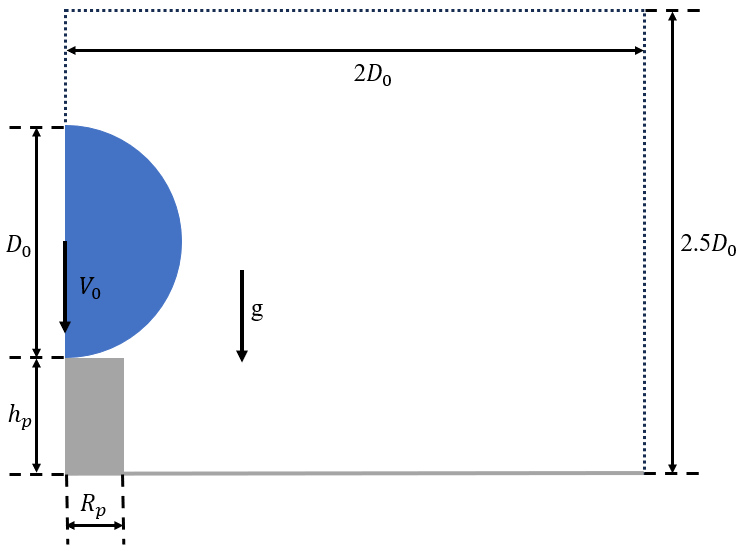}
\caption{Schematic depiction of the calculation domain}
\label{fig:Geo}
\end{figure}
In the present study, we focus on the droplet impact onto a cylindrical pillar at high velocities, wherein the compressibility effects cannot be neglected. The above mentioned governing equations are solved within a 3D-wedge domain with one cell thickness. The geometry of the current study is depicted in Figure~\ref{fig:Geo}. 

In the simulation, we consider a water droplet with a diameter $D_0=2$~mm, which impacts onto the center of a cylindrical pillar at a velocity of $V_0=100$~m/s. Initial conditions are established with a pressure of $0.1$~MPa and a temperature of $300$~K. Consequently, the Reynolds and Weber numbers are given by $Re=\rho_l V_0 D_0/\mu_l=224,045$ and $We=\rho_l V_0^2 D_0/\sigma=285,714$, respectively. The pillar has a height of $h_p=1$~mm. Three pillar radius $R_p=0.25$~mm, $R_p=0.5$~mm and $R_p=0.75$~mm were studied. Since the Reynolds number and the Weber number at the contact edge in the present conditions is very high, the contact angle effect is insignificant at initial impact stage. Therefore, a statistical contact angle with $90^{\circ}$ has been employed.

In terms of simulation configuration, symmetry conditions are applied to the front and back boundaries. The wall boundaries adhere to no-slip conditions and are adiabatic. All other boundaries are set to be continuous, allowing for the flow to either enter or exit the computational domain. Three mesh resolutions with grid sizes of $3.75\ \mu$m, $2\ \mu$m and $1\ \mu$m are simulated to perform a grid sensitivity study. Overall, the parameters of cases used in the present study is listed in Tab.~\ref{Tab:Cases}.
\begin{table}[h!]
\centering
\caption{Parameters used in the present simulations}
\begin{tabular}{ |c|c|c|c| }
 \hline
 Cases&A&B&C\\
 \hline
  Impact Velocity $V_0$ (m/s)&100&100&100\\
  \hline
  Droplet diameter $D_0$ (mm)&2&2&2\\
  \hline
  Reynolds number $Re$ &224,045&224,045&224,045\\
  \hline
  Weber number $We$&285,714&285,714&285,714\\
  \hline
  Radius of the cylindrical pillar $R_p$ (mm)&0.25&0.5&0.75\\
  \hline
  Height of the cylindrical pillar $h_p$ (mm)&1&1&1\\
  \hline
\end{tabular}
\label{Tab:Cases}
\end{table}

\subsection{Grid sensitivity study}
\begin{figure}[h]
\centering
    \begin{subfigure}[b]{0.48\textwidth}
        \centering
        \includegraphics[scale=0.47]{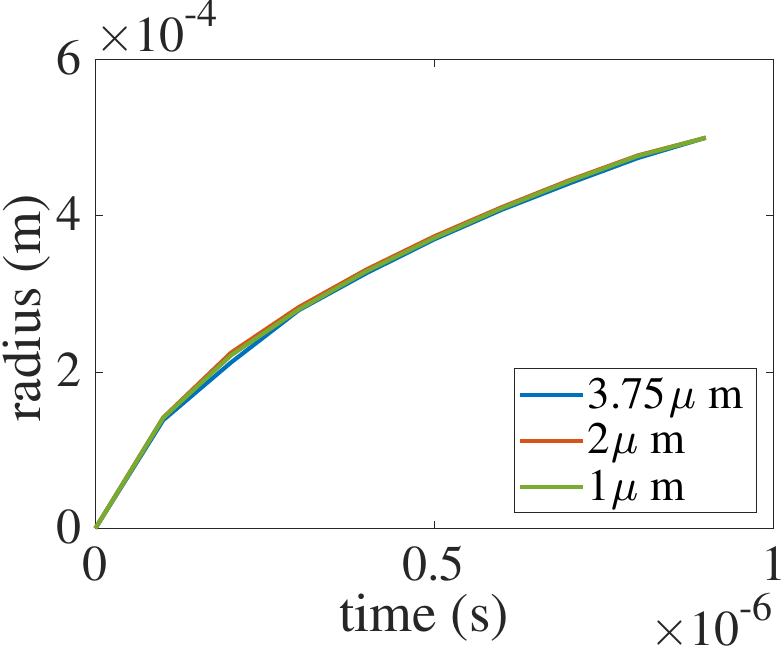}
        \caption{}
        \label{fig:Re_Gss}
    \end{subfigure}
    \begin{subfigure}[b]{0.48\textwidth}
        \centering
        \includegraphics[scale=0.47]{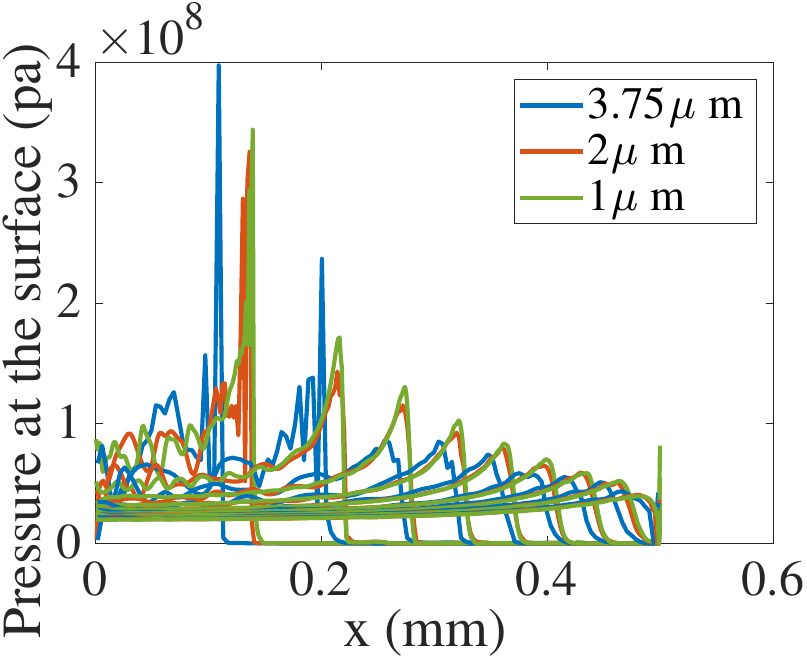}
        \caption{}
        \label{fig:P_Gss}
    \end{subfigure}
\caption{Radius of the turning line (a) and pressure evolution at the contact surface (b) under three grid resolutions.}
\end{figure}

The evolution of the radius of the droplet turning line, which will be defined in the subsequent section, is depicted over time for three distinct grid resolutions in Fig.~\ref{fig:Re_Gss}. The results obtained from the three mesh configurations are closely aligned, with only minor deviations observed between them. The pressure evolution over time at the contact surface of the droplet for the three grid resolutions is plotted in Fig.~\ref{fig:P_Gss}. It is demonstrated that simulations on the meshes with the resolutions of 2~$\mu$m and 1~$\mu$m predict comparable pressure distributions, while the simulation on the mesh with a resolution of 3.75~$\mu$m tends to overestimate the pressure at the initial stage of the impact and underestimates the pressure at later stage of the impact. Considering the computational cost and the accuracy of the simulation, the mesh with a resolution of 2~$\mu$m is used in the present study.

\section{Results and discussion}
\subsection{High velocity droplet impact}
\label{sec:dynamic}

\begin{figure}[h]
\centering
\includegraphics[scale=0.582]{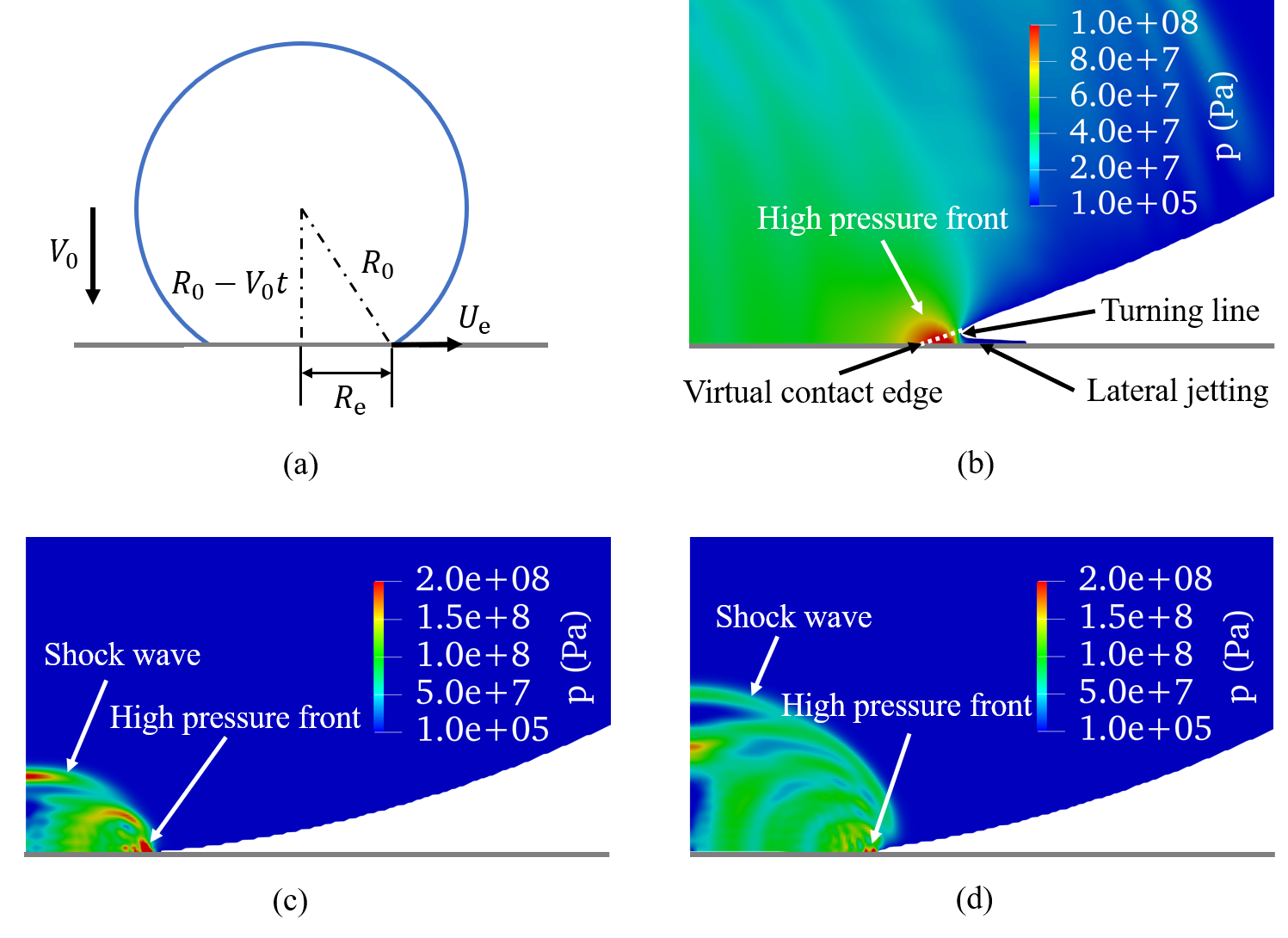}
\caption{Evolution of the pressure field near the wall: (a) Schematic depiction of the contact edge velocity; (b) Formation of the turning line and lateral jetting; (c) Shock wave attached with the contact edge; (d) Shock wave detaching from the contact edge.}
\label{fig:shock}
\end{figure}

Prior to presenting the results from the droplet-pillar impact, it is essential to discuss the liquid dynamics associated with a high-velocity droplet impacting a dry wall. When a droplet impacts onto a dry surface, the contact edge starts from a singular point. As illustrated in Fig.~\ref{fig:shock}a, the contact edge's radius can be analytically calculated as
\begin{equation}
\label{eq:Re}
    R_e = \sqrt{R_0^2-(R_0-V_0t)^2} = \sqrt{2R_0 V_0 t - V_0^2 t^2}
\end{equation}
The velocity of the contact edge is derived from taking the derivative of Eq.~(\ref{eq:Re}) by
\begin{equation}
\label{eq:Ue}
    U_{e}=\frac{V_0(R_0-V_0 t)}{\sqrt{2R_0 V_0 t-V_0^2t^2}},
\end{equation}
where $R_0$ represents the droplet's initial radius. An observable singularity at $t=0$ implies that the velocity of the contact edge is initially infinite and diminishes as the edge progresses outward. This high initial velocity compresses the liquid behind the contact edge, leading to a high-pressure front as shown in Fig.~\ref{fig:shock}. For low-velocity impacts, the time scale of the compressibility effect is small enough to be neglected. However, for impact velocity exceeding \(60~\text{m/s}\), the compressed liquid emits significant compression waves. As the contact edge moves outwards, the envelop of the compression waves generate a shock wave inside the droplet, which is illustrated in Fig.~\ref{fig:shock}(c-d). Initially, due to the high velocity of the contact edge, the shock wave is attached with the edge, as shown in Fig.~\ref{fig:shock}c. And the pressure at the contact edge increases. Once velocity of the contact edge diminishes below the shock wave's speed, the shock wave is detached from the contact edge and travels inside the droplet, as demonstrated in Fig.~\ref{fig:shock}d. The highly compressed liquid behind the contact edge begins to expand in the lateral direction, leading to a high-velocity lateral jetting. The pressure behind the contact edge begins to decrease correspondingly. As the lateral jetting develops, the connection of the jetting and the droplet is defined as the turning line, as highlighted in Fig.~\ref{fig:shock}b. This line's radius can be calculated by the minimal radius of the liquid gas interface near the wall. Moreover, the peak pressure at the contact surface is positioned immediately behind this turning line.

\begin{figure}[h]
\centering
\includegraphics[scale=0.5]{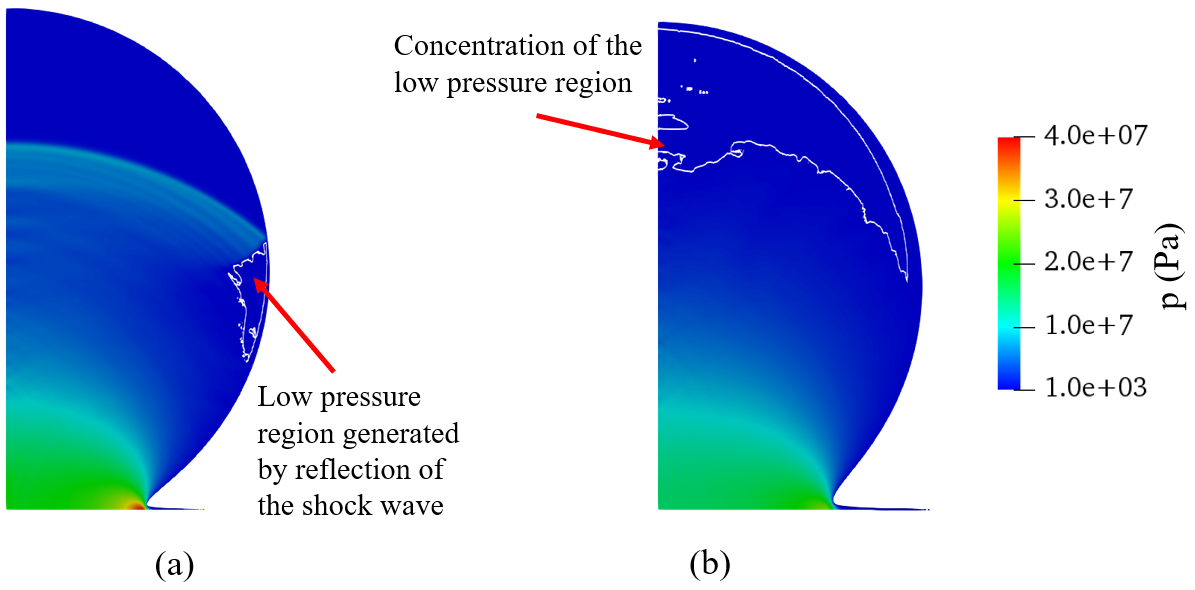}
\caption{Pressure field evolution within the droplet. White lines delineate the iso-lines with a value of \(0.001~\text{MPa}\): (a) Low pressure region generated by the reflection of the shock wave; (b) Concentration of the low pressure region at the center of the droplet.}
\label{fig:pressurePlane}
\end{figure}

As the shock wave propagates within the droplet, it reflects at the liquid gas interface as a rarefaction wave. A low pressure region is generated behind the shock wave near the interface. When the shock wave gets reflected from the top of the droplet, nearby low-pressure regions coalesce at the droplet's center. The concentration and superposition of the rarefaction waves lead to cavitation inside the droplet. The collapsing of the cavitation induces another shock wave, which is responsible for the erosion of the surface material. 

\subsection{Droplet impact onto a cylindrical pillar}
\begin{figure}[h]
\centering
\includegraphics[scale=0.5]{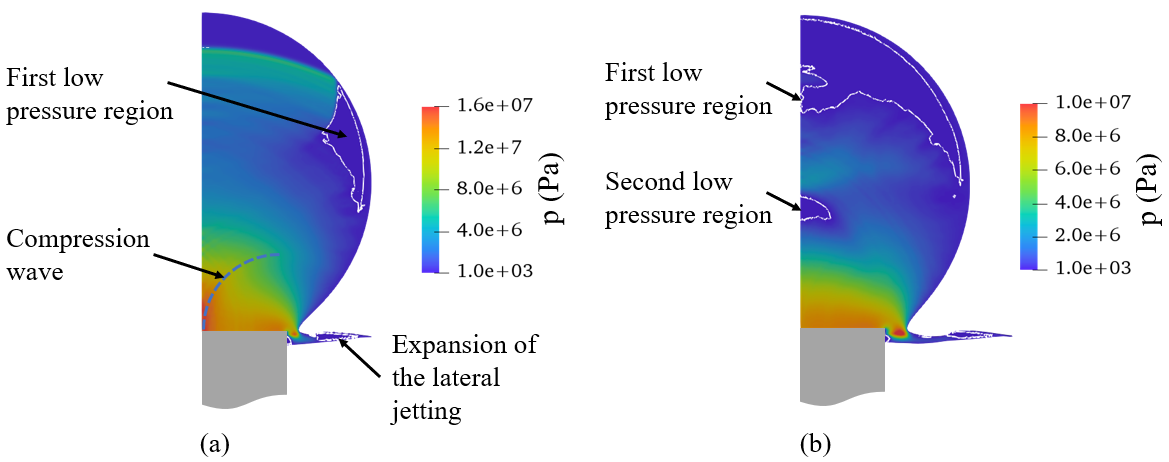}
\caption{Case B. (a) Compression wave generated by expansion of the high pressure front at the pillar edge and the first low pressure region near the interface; (b) Concentration of the first and the second low pressure regions.}
\label{fig:caseB}
\end{figure}
\begin{figure}[h]
\centering
    \begin{subfigure}[b]{1.0\textwidth}
        \centering
        \includegraphics[scale=0.5]{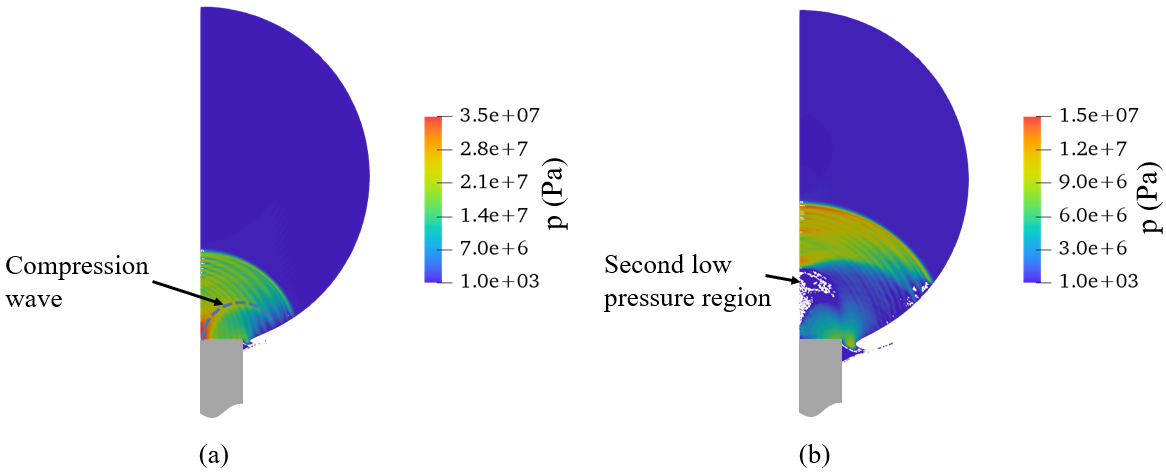}
    \end{subfigure}
    \begin{subfigure}[b]{1.0\textwidth}
        \centering
        \includegraphics[scale=0.49]{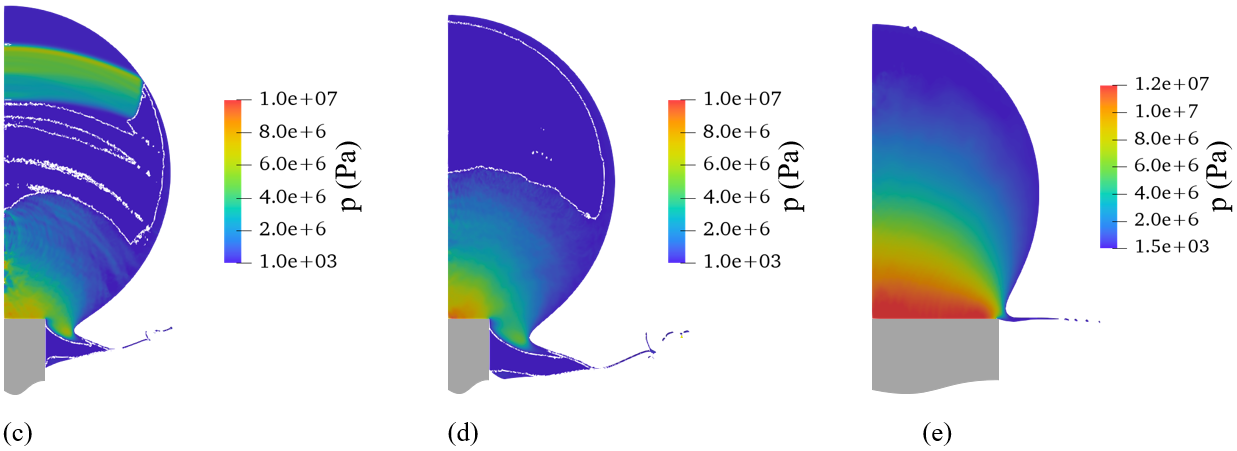}
    \end{subfigure}
\caption{Case A and Case C. (a) Compression wave generated by expansion of the high pressure front at the pillar edge for case A; (b) Generation of the first low pressure region for case A; (c) Merging of the first and the second low pressure region for case A; (d) Possible stronger cavitation for case A; (e) No generation of the compression wave by expansion of the high pressure front at the pillar edge for case C.}
\label{fig:caseAC}
\end{figure}
In this section, we discuss the evolution of pressure during an impact onto a cylindrical pillar. Figures~\ref{fig:caseB} and~\ref{fig:caseAC} depict iso-lines, representing a pressure iso-value of 0.001~MPa. As shown in Fig.\ref{fig:caseB}a, the compressed lateral jetting expands at the pillar edge rapidly. This high-pressure area is consistent with an impact on a flat surface until the turning line reaches the pillar edge. Figures~\ref{fig:caseB}a and~\ref{fig:caseAC}a show that the rapid expansion of highly compressed liquid releases a compression wave, subsequently followed by a rarefaction wave. This pattern is analogous to the Friedlander waveform \citep{Friedlander1946}. As the rarefaction wave moves toward the droplet's center, a second low pressure region is generated, as demonstrated in Fig.\ref{fig:caseB}b and Fig.\ref{fig:caseAC}b. For case~A, the expansion-induced low-pressure region coalesces with the shock wave reflection-induced low-pressure area, as seen in Fig.\ref{fig:caseAC}c. When this low-pressure region concentrates in the droplet's center, the existence of the additional rarefaction wave induced by expanded liquid at the pillar edge may lead to a stronger cavitation, as depicted in Fig.\ref{fig:caseAC}d. For case~C, this secondary rarefaction wave is absent, as illustrated in Fig.\ref{fig:caseAC}e. This absence can be attributed to the observed phenomenon in Fig.\ref{fig:P_Gss} where the pressure behind the turning line diminishes as the turning line travels outward. Thus, the expansion of the compressed liquid is not strong enough to generate a rarefaction wave.

\begin{figure}[h]
\centering
\includegraphics[scale=0.6]{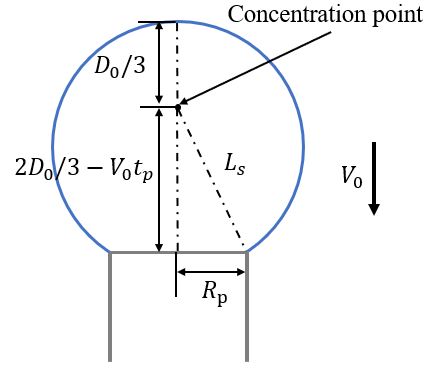}
\caption{Geometric depiction of droplet impact onto a cylindrical pillar: when the droplet edge arrives at the pillar edge.}
\label{fig:pillarGeo}
\end{figure}
From the above analysis, it becomes evident that there exists a critical pillar diameter at which the primary low-pressure region merges with the secondary one. This diameter correlates with the high-pressure region's growth immediately behind the turning line. Let $t_p$ represent the time it takes for the droplet to contact the surface and for the high-pressure front near the turning line to reach the pillar edge. Let $t_s$ indicate the time required for the second rarefaction wave to reach the cavitation point, and $t_c$ denotes the time needed for the reflected rarefaction wave to converge at the droplet's center. The condition for both low-pressure regions to meet at the concentration point is given by:
\begin{equation}
\label{eq:t-condi}
    t_p+t_s<t_c.
\end{equation}
The concentration point is located $D_0/3$ from the top of the droplet, as proposed by \citet{wu_xiang_wang_2018}, and is illustrated in Fig.~\ref{fig:pillarGeo}. Assuming that the shock propagation velocity equals the speed of sound, denoted as $c$, the time $t_c$ can be expressed as:
\begin{equation}
\label{eq:tc}
    t_c=4D_0/3c.
\end{equation}
As is illustrated in Fig.~\ref{fig:pillarGeo}, $t_s$ can be determined by:
\begin{equation}
\label{eq:ts}
    t_s = \frac{L_s}{c} =\frac{\sqrt{(2D_0/3-V_0 t_p)^2+R_p^2}}{c}.
\end{equation}
To predict the critical pillar diameter $R_{p,c}$, it is necessary to model the turning line's evolution $R_t$, as the highest pressure on the contact surface is located immediately behind the turning line.

\subsection{Modelling of the turning line and the high pressure front}
For incompressible flow, the turning line is modeled using the potential flow assumption combined with the Wagner condition \cite{wagner1932} and the self-similar approach by \citet{philippi_lagrée_antkowiak_2016}:
\begin{equation}
\label{eq:Rt-incomp}
    R_t = \frac{\sqrt{6D_0 V_0 t}}{2}.
\end{equation}
However, as will be demonstrated later, this model exhibits significant deviations when predicting high-velocity droplet impacts. To the author's knowledge, a model predicting the turning line radius and pressure behind it for compressible flow hasn't been proposed yet. In this section, such a model will be introduced.

For model development, the impact of a droplet on a rigid dry wall was simulated with two droplet diameters and five impact velocities, using water as the liquid. The cases employed for modeling are listed in Tab~\ref{Tab:MD_Cases}, with the impact Mach number defined as $\mathrm{Ma}=V_0/c$.

\begin{table}[h!]
\centering
\caption{Cases used in the modeling}
\begin{tabular}{ |c|c|c|c|c|c|c|c|c|c|c|c|c| }
 \hline
 Cases&1&2&3&4&5&6&7&8&9&10\\
 \hline
  $V_0$ (m/s)&60&100&150&200&250&60&100&150&200&250\\
  \hline
  $D_0$ (mm)&2&2&2&2&2&1&1&1&1&1\\
  \hline
  $Ma$ &0.040&0.067&0.101&0.134&0.168&0.040&0.067&0.101&0.134&0.168\\
  \hline
  Case name &D2V60&D2V100&D2V150&D2V200&D2V250&D1V60&D1V100&D1V150&D1V200&D1V250\\
  \hline
\end{tabular}
\label{Tab:MD_Cases}
\end{table}

\begin{figure}[h]
\centering
    \begin{subfigure}[b]{0.49\textwidth}
        \centering
        \includegraphics[scale=0.46]{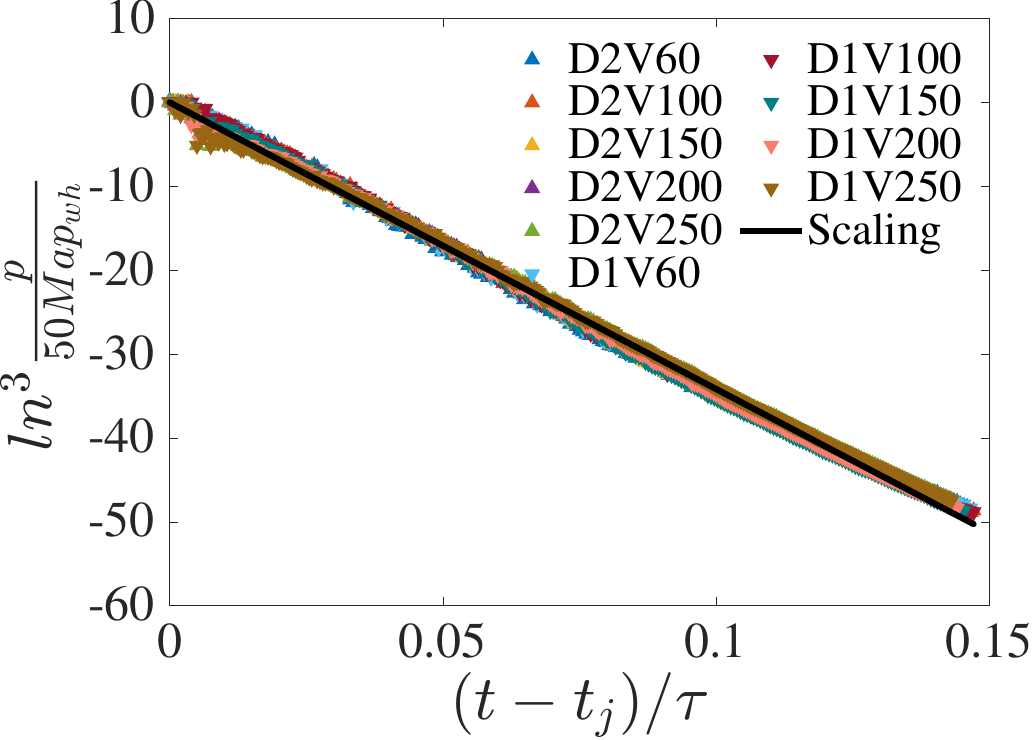}
        \caption{}
        \label{fig:P_scaling}
    \end{subfigure}
    \begin{subfigure}[b]{0.49\textwidth}
        \centering
        \includegraphics[scale=0.46]{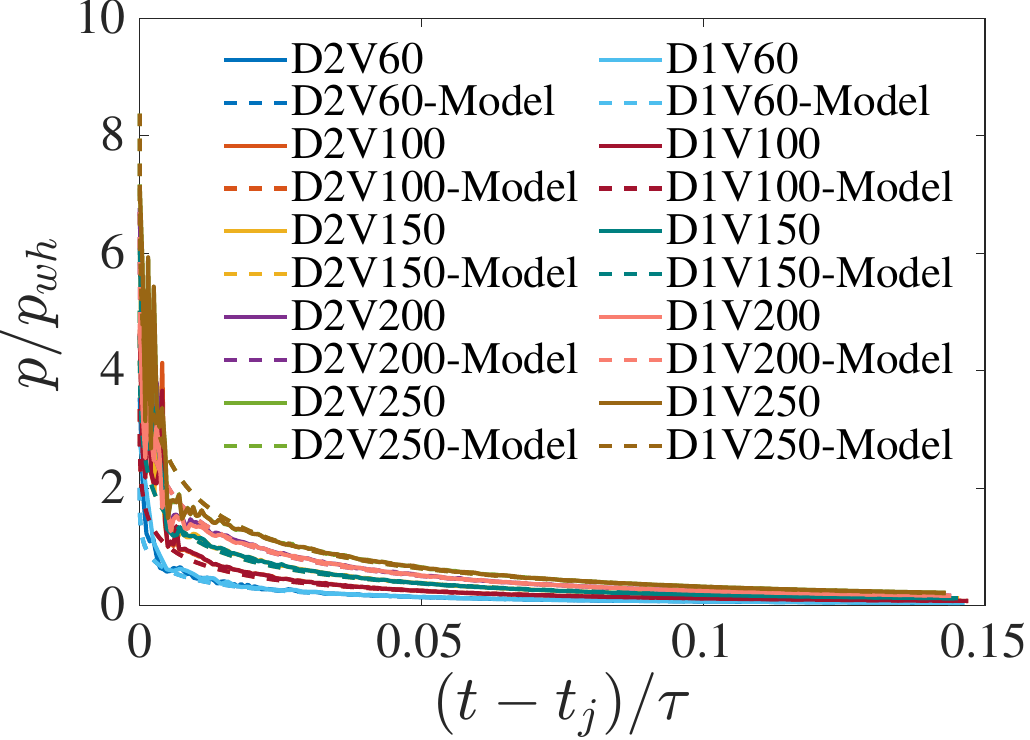}
        \caption{}
        \label{fig:P_plot}
    \end{subfigure}
\caption{Evolution of the maximum pressure in the contact region over time (a) Linear scaling of the pressure (b) Non-dimensionalized pressure and model prediction.}
\end{figure}

As explained in Sec.~\ref{sec:dynamic}, the turning line is formed by the ejection of the lateral jetting. To model the radius of the turning line, the characteristic length scale of the lateral jetting, $\delta$, is introduced in Eq.~(\ref{eq:Re}):
\begin{equation}
\label{eq:rt-Vj-pre}
    R_t = \sqrt{2R_0 V_0 t - V_0^2 t^2 +D_0 \delta}.
\end{equation}
The length scale $\delta$ is initially 0 and increases with the progression of the lateral jetting. Therefore, it is reasonable to assume that $\delta$ is proportional to the integration of the additional lateral jetting velocity $V_j$
\begin{equation}
\label{eq:delta}
    \delta \propto \int_{t_j}^{t} V_j(t) dt,
\end{equation}
where $t_j$ denotes the time when the shock wave detaches from the contact line. We adopt the model proposed by \citet{Haller2002} to estimate $t_j$
\begin{equation}
    t_j = \frac{R_0 V_0}{2\hat{s}^2},
\end{equation}
where $\hat{s}$ is the local shock velocity calculated by
\begin{equation}
    \hat{s} = s_0 +k V_0.
\end{equation}
$s_0$ and $k$ are two parameters depending on the liquid properties. For water, experimental data yield $s_0=1647$~m/s and $k=1.921$. As is mentioned in Section~\ref{sec:dynamic}, the driving force of the lateral jetting is the pressure difference between the high pressure front behind the turning line and the ambient. Therefore, the pressure behind the turning line and the additional lateral jetting velocity are highly correlated. It is assumed that the expansion process of the lateral jetting is isentropic and in equilibrium, which simplifies the problem. To calculate the additional velocity of the lateral jetting, the energy equation along an adiabatic stream tube is used:
\begin{equation}
\label{eq:equil-energy}
    h_j+ \frac{V_j^2}{2}= h_{j,m} + \frac{V_{j,m}^2}{2},
\end{equation}
where the subscript '$m$' denotes the state at the onset of lateral jetting, indicating that $V_{j,m}=0$~m/s. In order to express the enthalpy as a function of pressure, the following procedure is applied. Firstly, using the stiffened gas equation of state given by \citet{LEMETAYER2004265}
\begin{equation}
    \rho = \frac{p+B}{c_v T (N-1)},
\end{equation}
the temperature is referred to the state at the initial condition as
\begin{equation}
\label{eq:T_ref}
    \frac{T}{T_0} = \frac{p+B}{p_0+B}\frac{\rho_0}{\rho},
\end{equation}
where $c_v$ is the specific heat at constant volume. Secondly, by integrating the isentropic equation of state Eq.~(\ref{eq:EOS}) and substituting it into Eq.~(\ref{eq:T_ref}), we obtain
\begin{equation}
\label{eq:T_ref_ise}
    \frac{T}{T_0} = \left(\frac{p+B}{p_0+B}\right)^{1-\frac{1}{N}}.
\end{equation}
Then, using the expression for the specific enthalpy $h=N c_v T +q$ and inserting Eq.~(\ref{eq:T_ref_ise}) into Eq.~(\ref{eq:equil-energy}), the correlation between the additional lateral jetting velocity and the pressure behind the turning line is
\begin{equation}
\label{eq:V-p_pre}
    V_j^2 = 2 N c_v T_0 \left[ \left( \frac{p_m+B}{p_0+B}\right)^{1-\frac{1}{N}} - \left( \frac{p+B}{p_0+B}\right)^{1-\frac{1}{N}} \right],
\end{equation}
where the constant $q$ is determined to ensure that the fluid's internal energy equals $e_0$ at a specified reference state, defined by $p_0$ and $T_0$. The calculation of $q$ is given by \citet{LEMETAYER2004265} as: 
\begin{equation}
    q = e_0 - \frac{p_0+N B}{p_0+B}c_v T_0.
\end{equation}
Finally, substituting the expression for the speed of sound,
\begin{equation}
    c^2 = \left( \frac{\partial p}{\partial \rho}\right)_s = N\frac{p_0+B}{\rho_0},
\end{equation}
into Eq.~(\ref{eq:V-p_pre}), the additional velocity of the lateral jetting can be calculated from the pressure behind the turning line as
\begin{equation}
\label{eq:V-p}
    V_j = c\sqrt{\frac{2}{N-1} \left[ \left( \frac{p_m+B}{p_0+B}\right)^{1-\frac{1}{N}} - \left( \frac{p+B}{p_0+B}\right)^{1-\frac{1}{N}} \right]}.
\end{equation}
In this expression, the maximum pressure behind the turning line $p_m$ and its time evolution $p$ are unknown and need to be modeled. In the present study, DNS data are used to scale the pressure. For the purpose of non-dimensionalizing the pressure, the water hammer pressure is defined as $p_{wh} = \rho_0 V_0 c$. The time is then non-dimensionalized by $\tau = D_0/V_0$. As shown in Fig.~\ref{fig:P_scaling}, the pressure can be scaled as a function of
\begin{equation}
\label{eq:p_scale}
    ln^3\frac{p}{50 \mathrm{Ma} p_{wh}} = -341.87\frac{t-t_j}{\tau}.
\end{equation}
The coefficients 50 and -341.87 are obtained by fitting using the least squares method. For $t=t_j$, the peak pressure can be defined as $p_m=50\mathrm{Ma}p_{wh}$.
By rearranging Eq.~(\ref{eq:p_scale}), the non-dimensionalized pressure can be expressed as
\begin{equation}
\label{eq:nonD-p}
    \frac{p}{p_{wh}} = 50\mathrm{Ma}e^{-(341.87\frac{t-t_j}{\tau})^{1/3}}.
\end{equation}
A plot of the non-dimensionalized pressure $p/p_{wh}$ against the non-dimensionalized time $(t-t_j)/\tau$ is provided in Fig.~\ref{fig:P_plot}. It is demonstrated that the non-dimensionalized evolution of the pressure depends on the impact velocity and remains independent of the radius of the droplet. The model predicted by Eq.~(\ref{eq:p_scale}) approximates the maximum pressure in the contact region well, especially when $(t-t_j)/\tau>0.02$.

\begin{figure}[h]
\centering
\begin{subfigure}[b]{0.49\textwidth}
        \centering
        \includegraphics[scale=0.46]{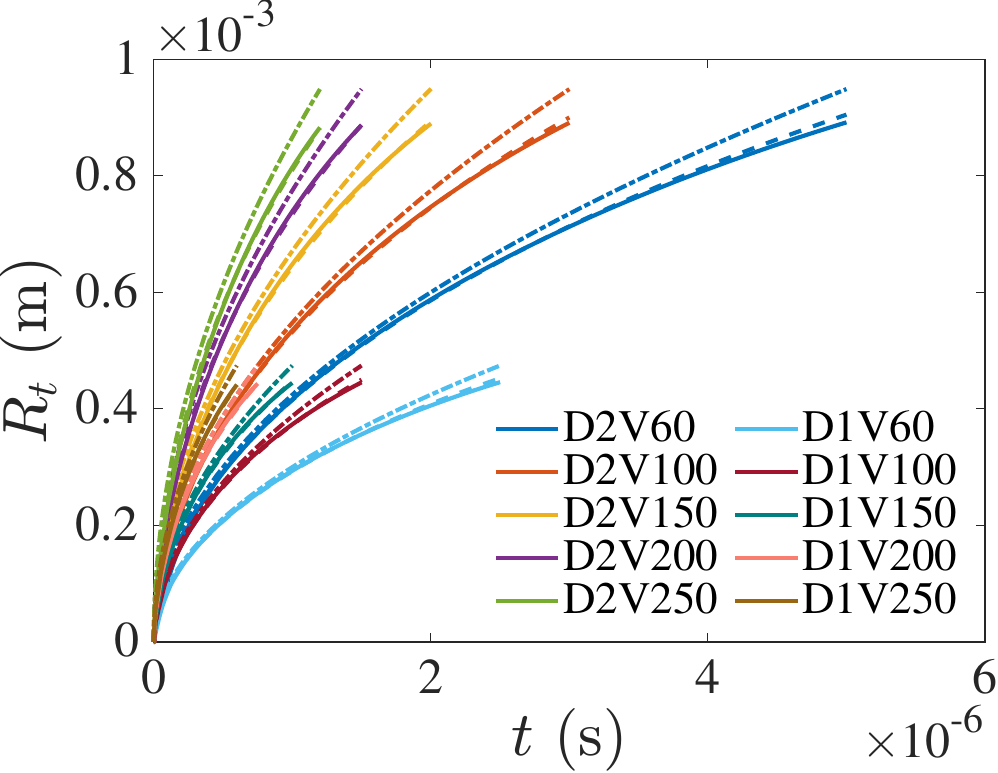}
    \end{subfigure}
    \begin{subfigure}[b]{0.49\textwidth}
        \centering
        \includegraphics[scale=0.46]{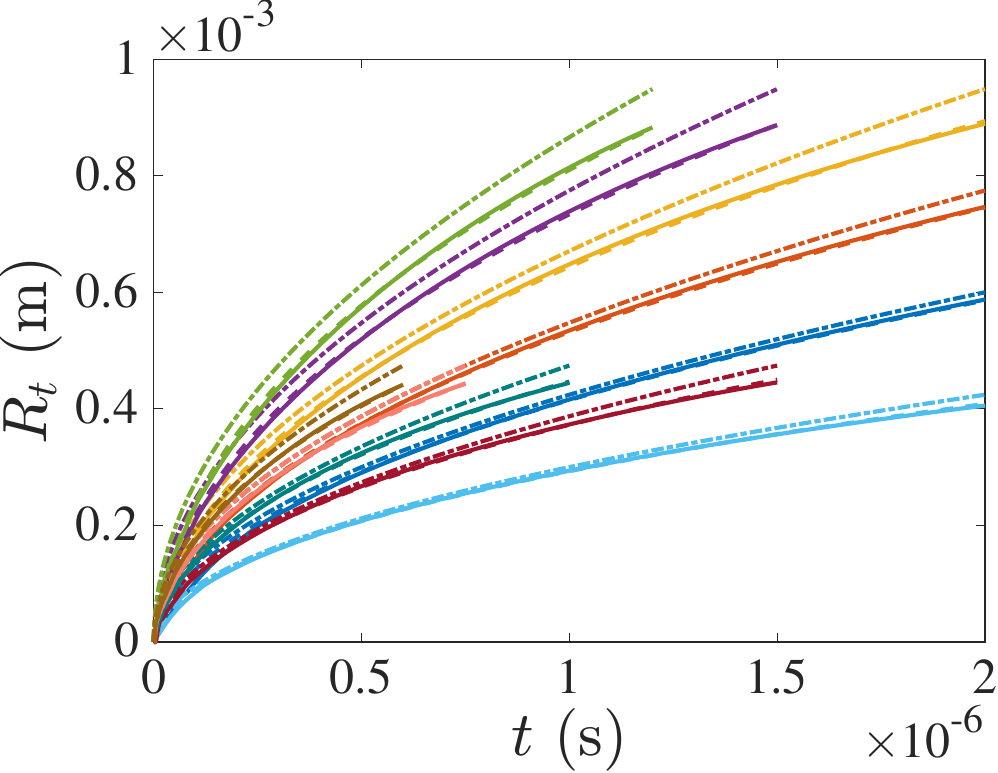}
    \end{subfigure}
\caption{Depiction of the radius of the turning line's evolution over time for all cases. Solid lines denote simulation data. Dashed lines represent the current model. Dash-dotted lines correspond to the incompressible model given by Eq.~(\ref{eq:Rt-incomp}). The figure on the right offers a closer view of the one on the left.}
\label{fig:Rt_model}
\end{figure}

Given the expression for pressure, the characteristic length scale of the lateral jetting can then be determined by substituting Eq.~(\ref{eq:V-p}) and Eq.~(\ref{eq:nonD-p}) into Eq.~(\ref{eq:rt-Vj-pre}). By introducing a coefficient $a$, which is derived from data calibration, the characteristic length scale can be expressed as
\begin{equation}
\label{eq:rt-exp}
    \delta = a \int_{t_j}^{t} c\sqrt{\frac{2}{N-1} \left[ \left( \frac{50\mathrm{Ma}p_{wh}+B}{p_0+B}\right)^{1-\frac{1}{N}} - \left( \frac{50\mathrm{Ma} p_{wh}e^{-(341.87\frac{t-t_j}{\tau})^{1/3}}+B}{p_0+B}\right)^{1-\frac{1}{N}} \right]} dt.
\end{equation}
On incorporating the above equation into Eq.~(\ref{eq:rt-Vj-pre}) and adjusting $a$ based on data, it is found that $a=0.055$. Figure~\ref{fig:Rt_model} provides a comparison between the data and the two models. It is evident that, the radius of the turning line predicted by the incompressible model, Eq.~(\ref{eq:Rt-incomp}) shows more significant deviations from the data as the impact velocity increases. On the contrary, the model proposed by the present study using Eq.~(\ref{eq:rt-Vj-pre}) and Eq.~(\ref{eq:rt-exp}) predicts accurately the radius of the turning line across all impact velocities and for two distinct droplet diameters.

Since the radius of the turning line has been accurately modeled, the time for the high pressure front near the turning line arriving at the pillar $t_p$ can be determined by using the following equation
\begin{equation}
\label{eq:rp-tp}
    R_p = \sqrt{2R_0 V_0 t_p - V_0^2 t_p^2 +D_0 \delta(t_p)}.
\end{equation}
The critical pillar radius $R_{p,c}$ is then obtained by solving the set of equations that includes Eq.~(\ref{eq:t-condi}), Eq.~(\ref{eq:tc}), Eq.~(\ref{eq:ts}) and Eq.~(\ref{eq:rp-tp}), using the Levenberg–Marquardt algorithm efficiently. For the present case with a droplet diameter of $D_0=2$~mm and an impact velocity of $V_0=100$~m/s, the computed critical pillar radius is $R_{p,c}=0.501$~mm. 

\section{Conclusion}

The fluid dynamics of liquid droplet impact on surfaces hold significant relevance to various industrial applications. In certain applications such as high-fogging system in gas turbines, steam turbines, flight vehicles through rain, medical inhaler with a liquid jetting nozzle and high-speed liquid jets in cleaning and cutting operations, the impact velocity of the droplet is relative high. In these applications, compressibility effects of the liquid cannot be neglected and is instrumental in causing material erosion. There is a lack of physical understanding as well as a proper analytical model. In the present study, we simulate a droplet impact onto cylindrical pillars of varying diameters and propose correlations for the evolution of the maximum impact pressure and the turning line of the droplet, aiming to predict pillar diameters in relation to possible cavitation enhancement. The liquid, which is highly compressed behind the contact edge, has been seen to expand sideways, resulting in rapid lateral jetting. As the droplet experiences the shock wave's progression, this wave is reflected as a rarefaction wave upon encountering the boundary between the liquid and the gas. This reflection creates a low-pressure zone right behind the shock wave, close to the interface. As the shock wave reflects off the droplet's peak, adjacent low-pressure areas merge at the center of the droplet. The merging and overlay of the rarefaction waves initiate cavitation within the droplet. The subsequent collapse of this cavitation generates another shock wave, which contributes to the surface material's erosion. Furthermore, an analytical model for the radius of the turning line is newly built. Contrary to the conventional incompressible model, which shows significant deviations with increasing impact velocities, the proposed model, closely mirrors the observed behaviors across a spectrum of velocities and droplet sizes. 

\section{Acknowledgments}

Yanchao Liu acknowledges the support by the Chinese Scholarship Council (CSC). All authors acknowledge the financial support by the German Science Foundation (DFG) under Germany's Excellence Strategy - EXC 2075 under the project number 390740016 and GRK 2160 under the project number 270852890. Guang Yang acknowledges the support by the National Natural Science Foundation of China (52276013). In addition, all authors gratefully acknowledge the access to the high performance computing facility Hawk at HLRS, Stuttgart.

\appendix
\section{Derivation of the volume fraction advection equation}
\label{sec:VoFeq}
For each phase, the mass balance equation can be expressed as:
\begin{equation}
\label{eq:contii}
    \frac{\partial (\alpha_i\rho_i)}{\partial t} + \nabla \cdot (\rho_i\alpha_i\mathbf{U}) = 0.
\end{equation}
By applying the product rule to Eq.~(\ref{eq:contii}), it can be reformulated as:
\begin{equation}
    \alpha_i\frac{\partial\rho_i}{\partial t}+\rho_i\frac{\partial \alpha_i}{\partial t} + \rho_i\alpha_i\nabla\cdot\mathbf{U}+\rho_i\mathbf{U}\cdot\nabla\alpha_i + \alpha_i\mathbf{U}\cdot\nabla\rho_i = 0.
\end{equation}
Rearranging this equation, one obtains:
\begin{equation}
\label{eq:rearrangedConti}
    \left( \frac{\partial\alpha_i}{\partial t} + \mathbf{U}\cdot\nabla\alpha_i \right) + \frac{\alpha_i}{\rho_i}\left( \frac{\partial \rho_i}{\partial t} + \mathbf{U}\cdot\nabla\rho_i \right) + \alpha_i\nabla\cdot\mathbf{U} = 0.
\end{equation}
By summing Eq.~(\ref{eq:rearrangedConti}) for both phases, the following expression is obtained:
\begin{equation}
\label{eq:addedConti}
    \left( \frac{\alpha_l}{\rho_l}\frac{D\rho_l}{Dt} + \frac{\alpha_g}{\rho_g}\frac{D\rho_g}{Dt} \right) + \nabla\cdot\mathbf{U} = 0.
\end{equation}
Substituting Eq.~(\ref{eq:addedConti}) into Eq.~(\ref{eq:rearrangedConti}) the continuity equation can be written as:
\begin{equation}
    \frac{\partial \alpha_l}{\partial t} + \nabla\cdot\left( \alpha_l \mathbf{U} \right) =
    \alpha_l\alpha_g\left( \frac{1}{\rho_g} \frac{\mathrm{D}\rho_g}{\mathrm{D} t} - \frac{1}{\rho_l} \frac{\mathrm{D}\rho_l}{\mathrm{D} t} \right) + \alpha_l \nabla \cdot \mathbf{U}.
\end{equation}
Typically, the volume fraction of the liquid phase $\alpha_l$ is denoted by $\alpha$ and the volume fraction of the gas phase is determined by $\alpha_g = 1- \alpha$. 

\bibliography{Refs}
\end{document}